# Ferroelectricity in the nematic liquid crystal under the direct current electric field

Mateusz Mrukiewicz[1]*, Paweł Perkowski[1], Jakub Karcz[2], Przemysław Kula[2]

[1]*Institute of Applied Physics, Military University of Technology, 2 Kaliskiego, 00-908 Warsaw, Poland*

[2]*Institute Chemistry, Military University of Technology, 2 Kaliskiego, 00-908 Warsaw, Poland*

*Corresponding author: mateusz.mrukiewicz@wat.edu.pl

ORCID numbers:

Mateusz Mrukiewicz: 0000-0002-0212-4520

Paweł Perkowski: 0000-0001-7960-2191

Jakub Karcz: 0000-0003-4783-0469

Przemysław Kula: 0000-0002-7862-7968

**Abstract**

We investigated the electrical properties of the liquid crystal compound 4-(4-nitrophenoxycarbonyl)phenyl 2,4-dimethoxybenzoate, known as an RM734, exhibiting a ferroelectric nematic phase. The influence of alternating (AC) and direct (DC) current electric fields on the switching process of the polarization vector and dielectric constant of planarly aligned ferronematic and nematic phases were examined. The decrease of the real part of electric permittivity in the ferronematic phase and the creation of ferroelectric order in the nematic phase under the DC field were demonstrated. The analysis of the results reveals the latching of the ferroelectric state. The applied DC field created a ferroelectric mode in the nematic phase. A new model of collective and molecular relaxations considering the domain structure of the ferronematic phase was proposed. The temperature and DC field dependence of dielectric properties were shown. The spontaneous polarization was measured using the field reversal technique. The spontaneous polarization value exhibits a maximum at the fixed temperature.

**Introduction**

Since the seventies, the well-investigated ferroelectric liquid crystal was the chiral smectic C phase (SmC*).[1] In the SmC* phase molecules are arranged in layers. In the layers, the long molecular axes of the molecules are parallel to each other. Furthermore, the long molecular axes are tilted at a fixed angle to the layer normal. The SmC* is a type of improper ferroelectric liquid material, where ferroelectricity comes from the coupling between polar and chiral orders, not from the concentration of dipole moments in the volume. Mayer predicted theoretically that by introducing chiral molecules into the SmC medium, we would eliminate the mirror symmetry operation in this structure.[2] The embedding of the chiral center in the rigid core of the molecule hinders the rotation of the molecule around its long axis. This implies a non-zero-sum of dipole moments in the direction perpendicular to the tilt plane.[3] For this reason, SmC* shows spontaneous polarization $\vec{P}$, where the vector $\vec{P}$ is parallel to the smectic layers. In SmC*, we do not need to have a sufficiently high concentration of dipolar molecules to create spontaneous polarization. Even one chiral molecule diluted in SmC matrix can create extremely low spontaneous polarization.[1] The spontaneous polarization in SmC* is rather low ($nC/cm^2$) compared to the common ferroelectrics. For the first time, the ferroelectric phase was experimentally observed in a material called DOBAMBC.[4] The understanding of the ferroelectricity state in the smectic liquid crystal resulted in the discovery of a new electro-optical effect, known as the Surface Stabilized Ferroelectric Liquid Crystal mode (SSFLC).[5] The ferroelectricity was also discovered in other layered ferroelectric liquid crystal phases by using bent-core molecules.[6–8]

In 1916 Max Born suggested that a phase transition from an isotropic phase to an anisotropic nematic liquid crystalline phase could be described as a transition from a paraelectric to a ferroelectric phase.[9] It was discovered later that the molecules which do not possess the permanent dipole moments can also create the nematic phase. Hence, the nematic phase is formulated due to the packing effect of the anisotropic molecules, and it has nothing to do with ferroelectricity. To observe molecular rearrangement between ordinary nematic and ferroelectric nematic we need highly polar mesogenic molecules to get a high enough concentration of dipoles to create a ferroelectric polar order, as in common ferroelectrics.[10–13] In the nematic liquid crystals in contrast to the smectic materials, the spatial distribution of the molecules shows no order, only one direction (along the director $\hat{n}$) is specified. The molecules do not lay in layers. According to the paper by Palffy-Muhoray, nematics formed by disk-shaped molecules with permanent dipole moments should form a ferroelectric nematic liquid crystal.[14] However, despite the theoretical research, for the first time, in 2017, a polar order was experimentally reported in two different non-chiral and highly polar molecules.[10,15] Mandle et al. presented the polar rod-like mesogenic material exhibiting a weak first-order phase transition between high- and low-temperature nematic phases, named N and $N_X$.[15,16] The ferroelectricity behavior of the $N_X$ phase, named in this paper as the ferronematic phase $N_F$, was confirmed by Chen et al..[17] The measured spontaneous polarization of $\sim 6 \, \mu C/cm^2$ was the largest ever reported for the ferroelectric fluid. Near the phase transition, the material exhibits a low value of the splay elastic constant $K_{22}$.[18] The other liquid crystalline material with confirmed ferroelectricity is a compound having a 1,3-dioxane unit in the mesogenic core (DIO) with a large dielectric anisotropy of more than $10^4$ and a high spontaneous polarization at the level of $\sim 4.4 \, \mu C/cm^2$.[10,19]

In the case of the phase transition between paraelectric SmA* and ferroelectric SmC* the dielectric constant measured at low frequencies regime increases ca. 5-50 times.[20–23] Such a significant increase is caused by the appearance of the Goldstone mode related to the presence of $P_S$. However, for the ferronematic-nematic phase transition, the situation is more complex. We present that one of the decisive factors that confirm the existence of the $N_F$ phase is spontaneous polarization and switching the polarization vector in the ferroelectric domains by the electric field. This work shows that a relatively high value of electric permittivity is not a determinant, which defines the existence of a ferroelectric nematic. Ordinary nematic built from highly polar molecules can exhibit a high value of electric permittivity comparable with the electric permittivity of the ferronematic phase, which is shown in this work.

In the studies of the ferroelectric nature of nematic liquid crystalline phase, the most important reference is the 4-[(4-nitrophenoxy)carbonyl]phenyl 2,4-dimethoxybenzoate, known as RM734.[15,16] From the application point of view, this material is of technological importance because of the presence of a wide temperature range in the ferroelectric nematic phase. So far, only a few compounds were reported to show the $N_F$ phase.[24,25] The research shows[17,26,27] that the ferronematic phase is very sensitive to the external electric field. However, the most of experiments were done using the AC field. Still, the electric properties of $N_F$ are not fully characterized and understood. Therefore, there is a need to provide precise dielectric spectroscopy measurements and descriptions of ferroelectric relaxations under a direct current electric (DC, bias) field. Here, we present dielectric properties, and dipole relaxation phenomena dependencies on the DC field and temperature. The whole results were compared and discussed. Furthermore, the full characterization of the mechanisms behind the formation of the $N_F$ phase including the switching of the polarization vector and temperature dependence of the spontaneous polarization was presented.

**Material**

The investigated material 4-(4-nitrophenoxycarbonyl)phenyl 2,4-dimethoxybenzoate, CAS Registry Number 2196195-87-4, the acronym RM734 is often used, was synthesized for the first time by Mandle et al..[15,16] To study the electrical and optical behavior of the ferronematic phase, we redesign the synthesis of the title compound using a different synthetic approach, where the opposite strategy of core formation was applied with special care of molecular purity determined as 99.8% (HPLC)[28]. Then we decided to study the prepared material the temperature range of the existence of nematic phases, optical switching, its dielectric behavior, and spontaneous polarization. The material exhibit, according to differential scanning calorimetry (DSC) in the cooling cycle at the scanning rate of 0.2 K/min, the following phase sequence: Iso (184.1°C) N (127.2°C) $N_F$ (< 0°C) Cr. In the cooling cycle, the calorimetric measurements show that the N-$N_F$ phase transition is accompanied by a relatively small enthalpy of ~0.50 J/g compared to ~1.00 J/g in the Iso-N phase transition, which indicates a slight change in the molecular arrangement between the phases. The N-$N_F$ transition does not change the orientation order but changes the directions of the dipole moments. The molecule is characterized by a large longitudinal dipole moment $\mu \cong 11$ D.[18] The dipole moment is directed along the long axis of the molecule which corresponds to the minimum of the momentum of inertia.

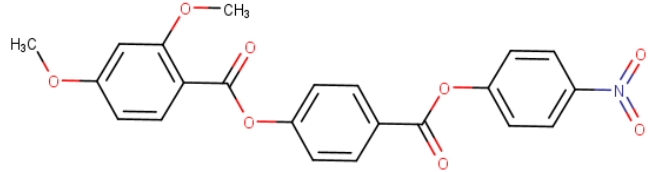

Figure 1 Chemical structure of RM734.

**Experiment**

Because of high $T_{N-I}$ the cell was filled with RM734 in the nematic phase (150°C) next rapidly heated to the isotropic phase (205°C) and then slowly cooled (0.3-0.5 K/min) to room temperature. The electro-optical studies were done at planarly aligned in-plane switching (IPS) cells where the molecular director $\hat{n}$ is parallel to the bounding surfaces. The electrodes were spin-coated with polyimide SE-130 (Nissan Chemicals). The rubbing process to obtain a uniformly planar alignment was anti-parallel along the electrodes. The width of the electrodes was 10 μm, while the distance between the electrodes in IPS cells was 40 μm. The picture of textures was taken by the polarized optical microscopy Jenapol (Carl Zeiss Jena) using a Delta optical camera (DLT-Cam Pro). The electric signal was generated using a Keysight 33500B waveform generator. The real and imaginary parts of electric permittivity were calculated using the following formulas:

$$\varepsilon'(f) = \frac{C}{C_0}, (1)$$

$$\varepsilon''(f) = \frac{G}{2\pi f C_0}, (2)$$

where $C_0$ is the capacitance of an empty cell. Capacitance ($C$) and conductance ($G$) in parallel equivalent circuit were determined by using an impedance analyzer Hewlett-Packard 4192A in the frequency ($f$) range of 100 Hz to 10 MHz. Dielectric spectroscopy (DS) studies were carried out in the planarly aligned 20 μm thick cells in form of parallel plate capacitors with gold electrodes of sheet resistance less than $1 \frac{\Omega}{\text{sq}}$. Its guarantees that dielectric spectroscopy is free of distortions up to a few MHz.[29] The homogenous (HG, planar) alignment of the liquid crystal molecules was achieved in the same way as for the IPS cell. The impedance analyzer generated a bias (DC) electric field from 0 to 10 V. The measuring field (AC) was 0.1 V. The temperature of the filled cells used for electro-optical and dielectric studies was stabilized by using a Linkam TMS 92 temperature controller with a heating stage Linkam THMSE 600, with an accuracy of 0.1 K.

**Results and discussion**

The phase transition between the nematic and ferronematic liquid crystal was studied at first by polarizing optical microscopy (POM). To observe well-crated ferroelectric domains slow cooling at the rate of 0.5°C/min was used. We noticed that when the cooling rate is too high the investigated material can crystallize instead of creating the $N_F$ phase. Moreover, even when a ferroelectric phase is formed, lowering the sample temperature quickly results in the observation of micrometer-scale domains instead of millimeter-scale domains.[26] During the cooling at the high rate polar molecules of RM734 do not have enough time to adapt their orientational and positional order to two thermodynamically stable states. However, during the slow cooling, molecules can easily change their positions. All experiments were done under isolated conditions to ensure the stability of the ferroelectric phase. In Figure 2 one can see that the isotropic-nematic phase transition behaves as typically as first-order (Figure 2a). The nematic phase is strongly and uniformly planarly aligned along the rubbing direction, which coincides with the $\hat{n}$ direction (Figure 2b). Below the nematic phase, subsequent temperature-dependent imaging reveals the N-$N_F$ phase transition. The transition begins with a variation of birefringence manifested by a gradual change in the color of the texture, appearing as green spots (Figure 2c). Next, the N texture roughens after cooling toward the ferronematic phase (Figure 2d). The used spacers become the source of the disclinations lines which lead to the creation of the domains (Figure 2e-f). Some of the domains are in the form of closed-lens-shaped domains while others are stretched along $\hat{n}$. As the temperature decreases, domains increase their area (Figure 2g). The fully created $N_F$ phase possesses oblong oval domains which are extended along the rubbing direction (Figure 2h). The visible disclination lines split areas of opposite polarization vectors. It is worth underlining that the ferroelectric domains are created spontaneously in the absence of the electric field. The spontaneous polarization for the whole sample is zero. The concentration of these domains is different in the cell bulk. The domains are stable for days. The creation process of the domains is repeated even after heating to the isotropic phase or cooling to the crystallization. From the temperature point of view, the whole N-N$_F$ transition process is very long, it needs a temperature change of around 20 degrees. No crystallization was observed at room temperature, although the texture changed significantly (Figure 2i).

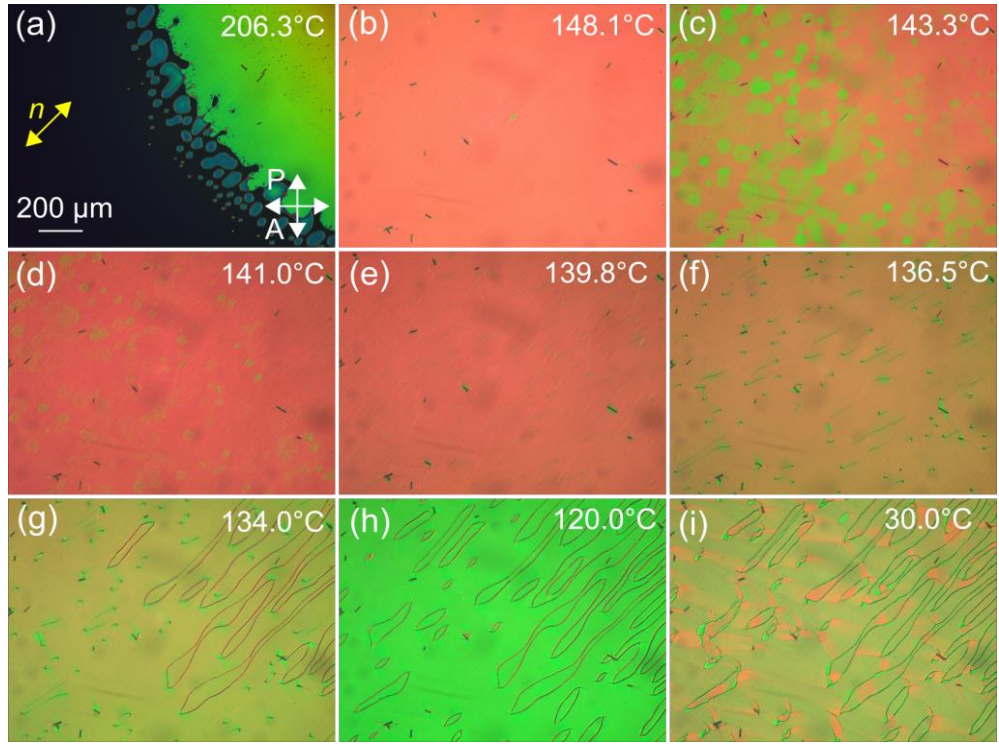

Figure 2 The phase transitions and mesophase textures of RM734. (a) Isotropic-nematic phase transition. (b) Planarly aligned nematic phase. (c-f) The phase transition process of creating the ferronematic phase. (g) Nematic-ferronematic phase transition. (h) High-temperature domain structure of the N$_F$ phase. (i) Room-temperature texture of RM734. In the IPS cell, the rubbing direction is parallel to the IPS electrodes.

In the absence of the electric field, we have uniform nematic alignment with the uniform orientation of the director $\hat{n}$ and with different orientations of the electric polarization vector $\vec{P}$ inside the neighboring domains.[17,30,31] In the ferronematic domains, due to the molecular structure of the investigated compound, the $\vec{P}$ vector is parallel to the molecular director $\hat{n}$, defined by the rubbing process (Figure 3a) and parallel to the IPS electrodes. Rotating the cell anti-clockwise (Figure 3b), where $\hat{n}$ is parallel to the polarizer direction, and clockwise (Figure 3c), where $\hat{n}$ is perpendicular to the polarizer direction showing no change of the texture color inside and outside domains. This means that there is no orientation of $\hat{n}$ between the domains. The director field inside the domains is uniform. The only change is visible on the boundaries of the domains where a 180° twist of $\hat{n}$ is observed. Therefore, at the boundaries the $\vec{P}$ vector transform to the opposite sign of $\vec{P}$. The blue and red arrows in Figure 3 indicate the polarization vector of the opposite sense. The resulting material polarization, without the applied electric field, is zero because the material tends to obtain minimum free energy.

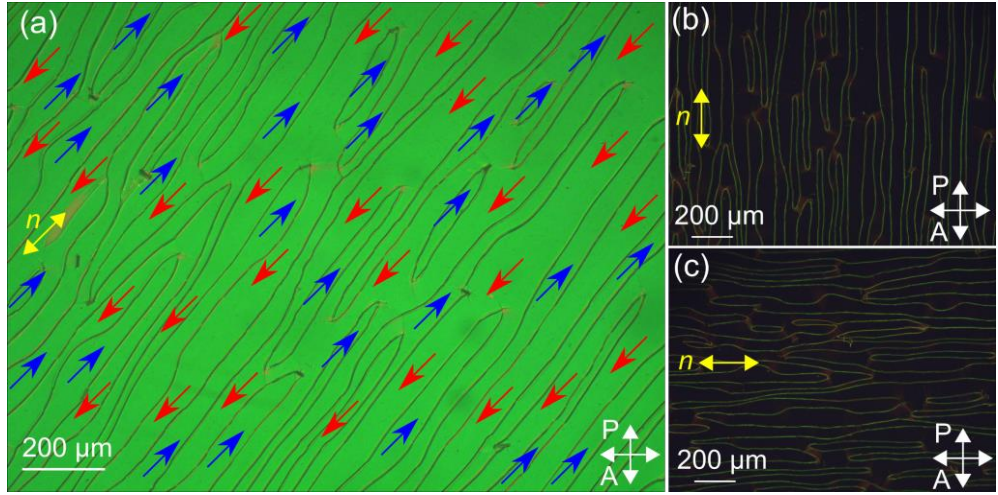

Figure 3 (a) Ferroelectric domains in the ferronematic phase. The blue and red arrows indicate domains with different orientations of the polarization vector $\vec{P}$. The $N_F$ texture picture after rotating the cell counterclockwise (b) and clockwise (c).

Figure 4 presents the step-by-step response of the ferroelectric domains to the applied AC electric field ($E_{AC}$) in the IPS cell. The measurements were done at 100°C ($T_{N-NF} = 136°C$), where the $N_F$ phase is fully created, and further cooling does not change the shape and size of the domains. To observe under the polarizing microscope the rotation of the polarization vector in the domains we applied the triangular wave (Figure 4a) in-plane electric field signal of low voltage ($U = 3$ Vpp) and low frequency ($f = 50$ mHz) to the planarly oriented cell. The width of the pulse was chosen to observe and deeply analyze the nucleation of domains with opposite polarity. After applying $E_{AC}$, we started observing the different electro-optical responses of the domains. As the voltage decreases from + 3Vpp to – 3Vpp the domains change their textural color. The polarization vector $\vec{P}$ begins to switch, toward the electric field $\vec{E}$, in one kind of domain in which the polarization coincides with the applied electric signal (Figure 4b). The domains with different polarities do change their structural color over time and maintain a fixed position. The polarization vector $\vec{P}$ is switched on both domains. All intermediate states (Figure 4b-f) are unstable in absence of the electric field. The fully reversible polarization switching process takes place without noticeable shrinking, moving, or annihilation of the domains. If the field is switched off the domains relax to the initial state with some relaxation time.

The switching using the DC electric field ($E_{DC} = \pm 2$ V/40 µm) looks similar to the switching using $E_{AC}$. Depending on the polarity of the $E_{DC}$ field we switch only one type of the domains (Figure 5a, b). The difference is that the application of a DC signal causes the formation of new domains over time with a polarity like the DC signal. Leaving a low amplitude $E_{DC}$ signal for a specific time will result in one type of domain (Figure 5c). After applying a low DC field in a well-aligned nematic phase (Figure 5d), we observed in the textures the formation of domains with a different director orientation than the rest of the sample. The green birefringence color is replaced in the domains with yellow, reflecting a higher orientational order (Figure 5e, f). After releasing the field, the domains which are created under the DC field disappear with time. While domains in the $N_F$ phase are preserved without an applied electric field. These domains with a higher orientational order can be created in the whole temperature range of the nematic phase. Of course, further increasing the field will cause the standard switching of the nematic order in the IPS cell.

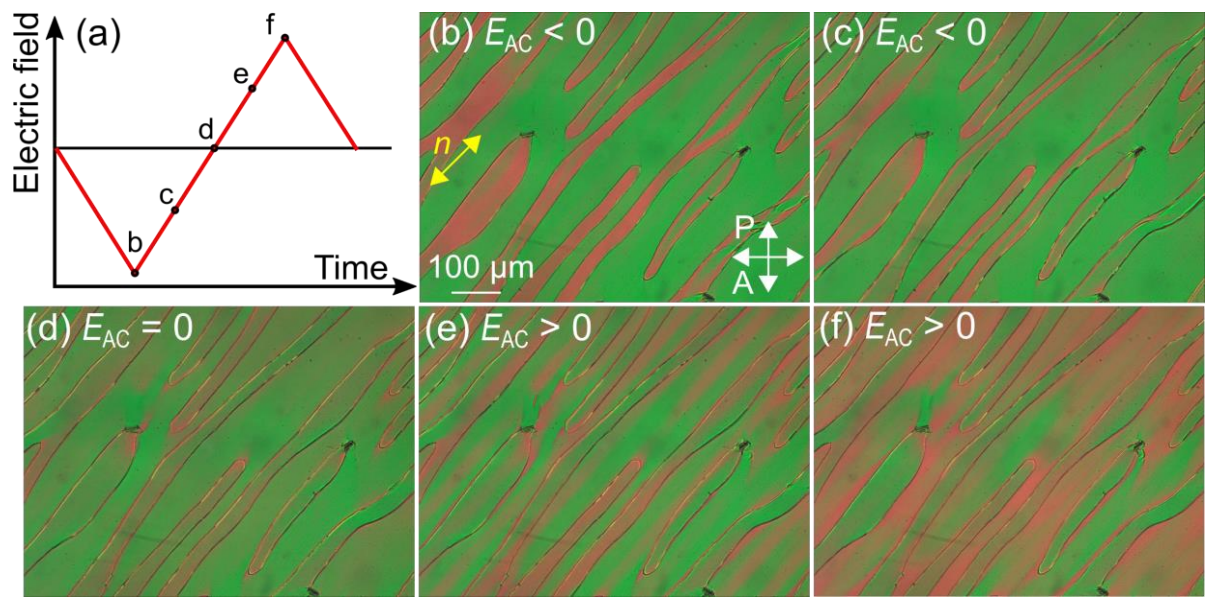

Figure 4 Rotation of the polarization vector in the ferronematic phase under AC electric signal. (a) Triangular in-plane wave electric signal. (b-f) The response of the ferroelectric domains with different polarities of the applied electric field.

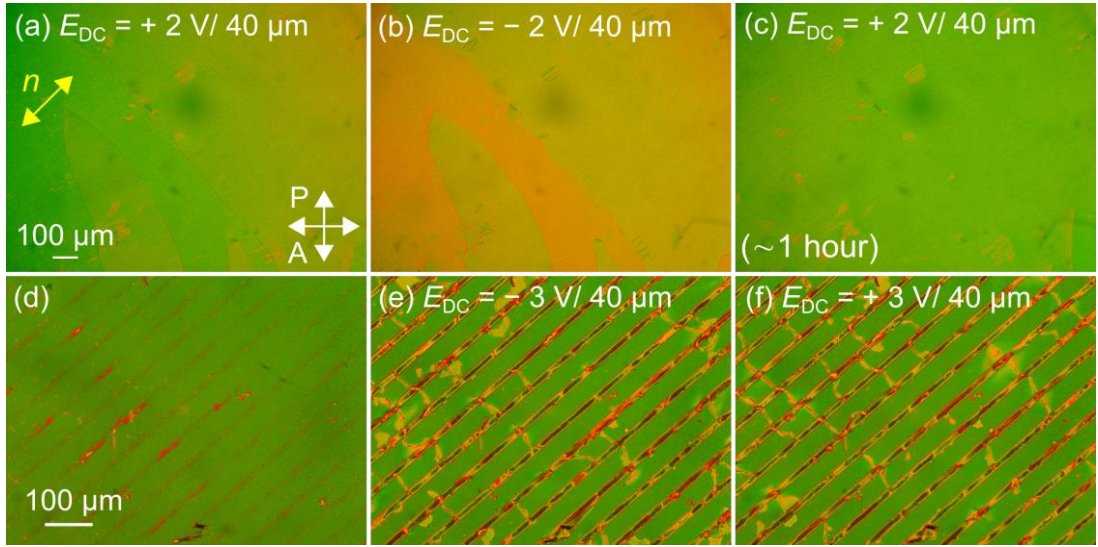

Figure 5 Electro-optical response of (a-c) the ferronematic phase at 100°C (d-f) the nematic phase at 160°C, after applying the DC field in the IPS cell.

To deeply analyze the influence of the bias electric field on electric permittivity values of the ferronematic phase, the dielectric spectroscopy measurements were done with bias fields at 0, 2, 4, 6, 8, and 10 V at 20 μm thick planarly aligned cell. Figures 6a and 6b present, in form of three-dimensional plots, the temperature, and frequency dependence of the real part of electric permittivity for measurements without (BIAS = 0 V) and with applied bias field (BIAS = 4 V). In measurements without the bias electric field, the dielectric spectrum shows the well-distinguishable transition between the nematic and ferronematic phases. The ferronematic phase is characterized by a much higher value of electric permittivity at low frequencies than the nematic phase. For instance, the value of $\varepsilon'$ in the $N_F$ phase equals to $\varepsilon' = 5488$ at $f = 1$ kHz and $T = 100°C$, while in the N phase is almost forty times lower, $\varepsilon' = 152$ at $f = 1$ kHz and $T = 160°C$. The measured values of the real part of electric permittivity decrease rapidly with the increase of the frequency of the measuring field (Figure 6c, d). The high values of $\varepsilon'$ in the ferronematic phase result from the presence of a ferroelectric mode or modes that do not occur in the nematic phase, and which are the subject of this article. We can assume that the dielectric modes occurring in the ferroelectric phase should have a different character than those occurring in the paraelectric phase. In the N phase, the highly polar molecules of RM734 make a significant contribution to the real part of electric permittivity but they do not form the ferroelectric phase. In the cooling cycle, as we approach the N-$N_F$ phase transition, the sharp increase of the electric permittivity is visible. The maximum of $\varepsilon'$ is observed at 116°C in the $N_F$ phase. Then, the dielectric constant slowly decreases with temperature decreasing to the value of 8 (at $T = 40°C$ and $f = 1$ kHz). After applying the bias electric field, we observed a significant decrease of $\varepsilon'$ in the ferronematic phase. As the DC field value increases, the decrease in permittivity becomes more and more visible (see Supplementary Material). The situation is much more complex in the nematic phase than in the ferronematic. Initially, the application of a low bias field (BIAS = 2 V) results in a marked increase in the real part of electric permittivity. However, this increase becomes smaller, but still noticeable for higher BIAS values. For high bias voltages of 8 and 10 V, the phase transition from N to $N_F$ becomes invisible. The phase transition becomes continuous, with no apparent sudden change in the value of $\varepsilon'$.

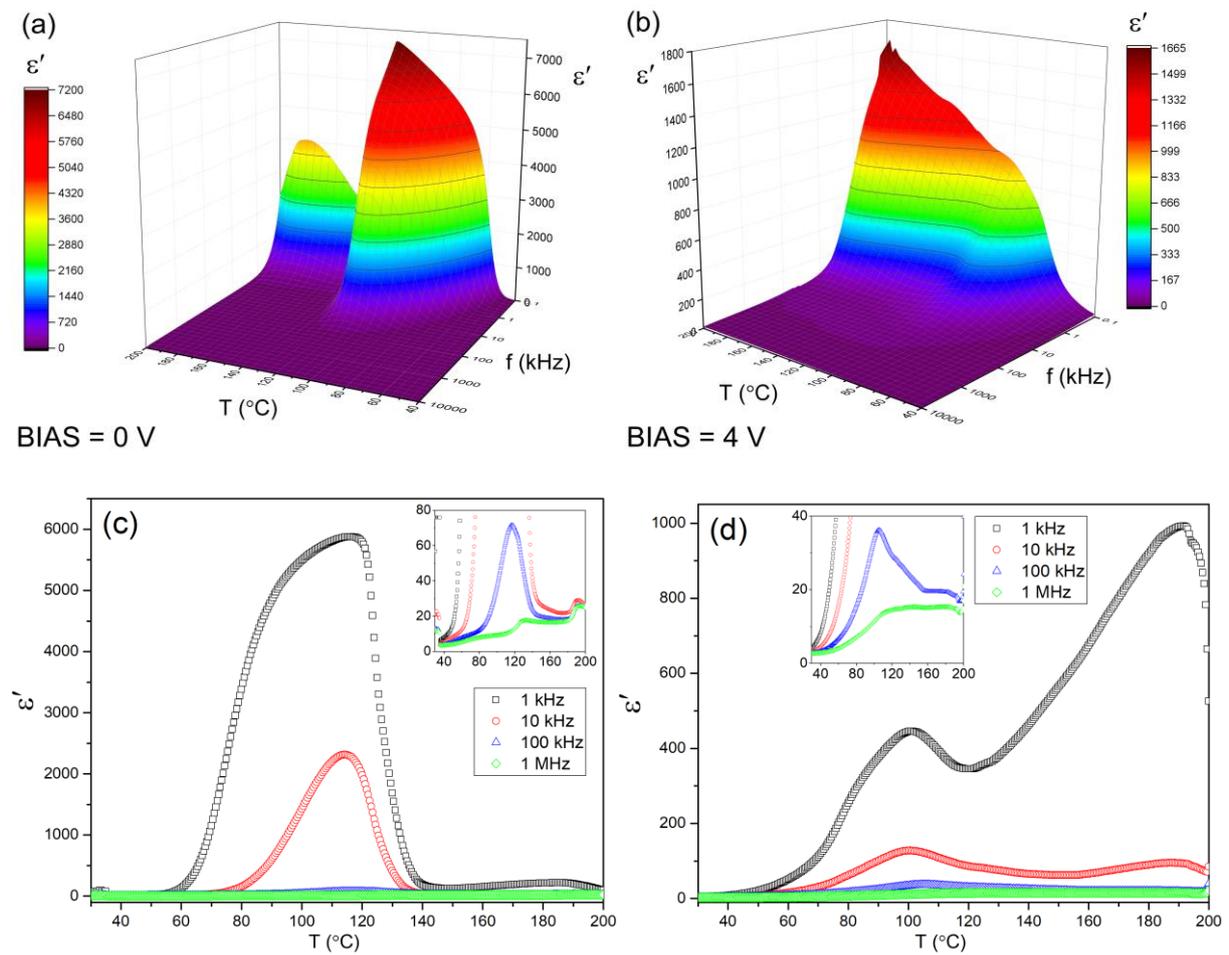

Figure 6 3D (a, b) and 2D (c, d) graphs of real part $\varepsilon'$ of electric permittivity versus frequency $f$ and temperature $T$, measured in planarly aligned cell of RM734 without (a, c) and with (b, d) the bias electric field.

To describe the changes in the permittivity values in the ferronematic phase and the nematic phase after applying the bias electric field, the experimental data were ($\varepsilon'$ and $\varepsilon''$) fitted to the Cole-Cole formula (the C-C model)[32] with the ionic contribution.[33] To know the relaxation time and amplitudes of the existing modes, we calculate the real $\varepsilon'_M$ and imaginary $\varepsilon''_M$ parts of electric permittivity from the C-C model in each phase:

$$\varepsilon^* = \varepsilon'_M - j\varepsilon''_M = \varepsilon_\infty + \sum_i \frac{\delta\varepsilon_i}{1+\left(j\frac{f}{f_{R_i}}\right)^{1-\alpha_i}} - j\frac{\sigma}{2\pi f \varepsilon_0}. \tag{3}$$

In the following formula $\varepsilon_\infty$ is the high-frequency limit of permittivity, $\varepsilon_0$ is the electric permittivity of free space, $\delta\varepsilon_i$ denotes the dielectric strength of $i^{th}$ dielectric mode, while $f_{R_i}$ is the relaxation frequency of $i^{th}$ dielectric mode. The distribution parameter $\alpha_i$ describes the relaxation spectra, for pure Debye relaxation $\alpha = 0$. The parameter $\sigma$ represents the contribution of the ionic conductivity at low-frequency measurements. Thanks to the C-C model we were able to determine the relaxation frequency (relaxation time) and dielectric strength of the ferroelectric and paraelectric modes and their correlation the to bias electric field. Comparing the temperature and DC-field dependence of relaxations parameters we can verify the molecular or collective nature of the dielectric modes in the ferronematic phase.

Figure 7 shows an example of complex permittivity versus the frequency (on a logarithmic scale) at two characteristic temperatures of the N and $N_F$ phase for measurements with and without the bias electric field. In all cases, the solid lines of the model are well-fitted to the full and empty black squares of the experimental data. The nematic phase without the DC field exhibits only one low-frequency relaxation process (Figure 7a). The distribution parameter is lower than 0.1 which means that the dielectric spectra are dominated by one dielectric mode with a strictly defined relaxation frequency. In the ferronematic phase (Figure 7b) we need three modes with the relaxation frequencies of 0.3, 4, and 9 kHz to fit the data. We believe that the nature of the low-frequency mode observed in the ferronematic phase is the same as the low-frequency mode in the nematic phase. Therefore, the existence of the ferronematic phase is related to the occurrence of the mid-frequency and high-frequency modes, named Ferro-A and Ferro-B in this paper. In the dielectric spectra, also the relaxation process related to the rotation around the molecular long axis, with a relaxation frequency higher than 10 MHz, is observed. However, this mode is of no interest to the authors. Figures 7c and d show the influence of the 4 V DC field on the dielectric behavior of the N and $N_F$ phase, respectively. The bias electric field creates additional relaxation in the nematic phase. Figure 7c shows the low-frequency relaxation, as before, and the new middle-frequency relaxation with the peak at 1.8 kHz. The DC-created mode (Ferro-C) is continuously detectable in the whole temperature range of the $N_F$ phase. We believe that the Ferro-C mode is connected to the ferroelectricity because its relaxation frequency is similar to the relaxation frequency of the ferroelectric modes A and B. The relatively high distribution parameter of DC-created ferroelectric mode ($\alpha = 0.4$) suggests that it is a combination of two ferroelectric modes Ferro-A and Ferro-B, suppressed by the DC field. Therefore, in measurements with the bias field, the low-frequency mode makes the biggest contribution to $\varepsilon'$ in the ferronematic phase.

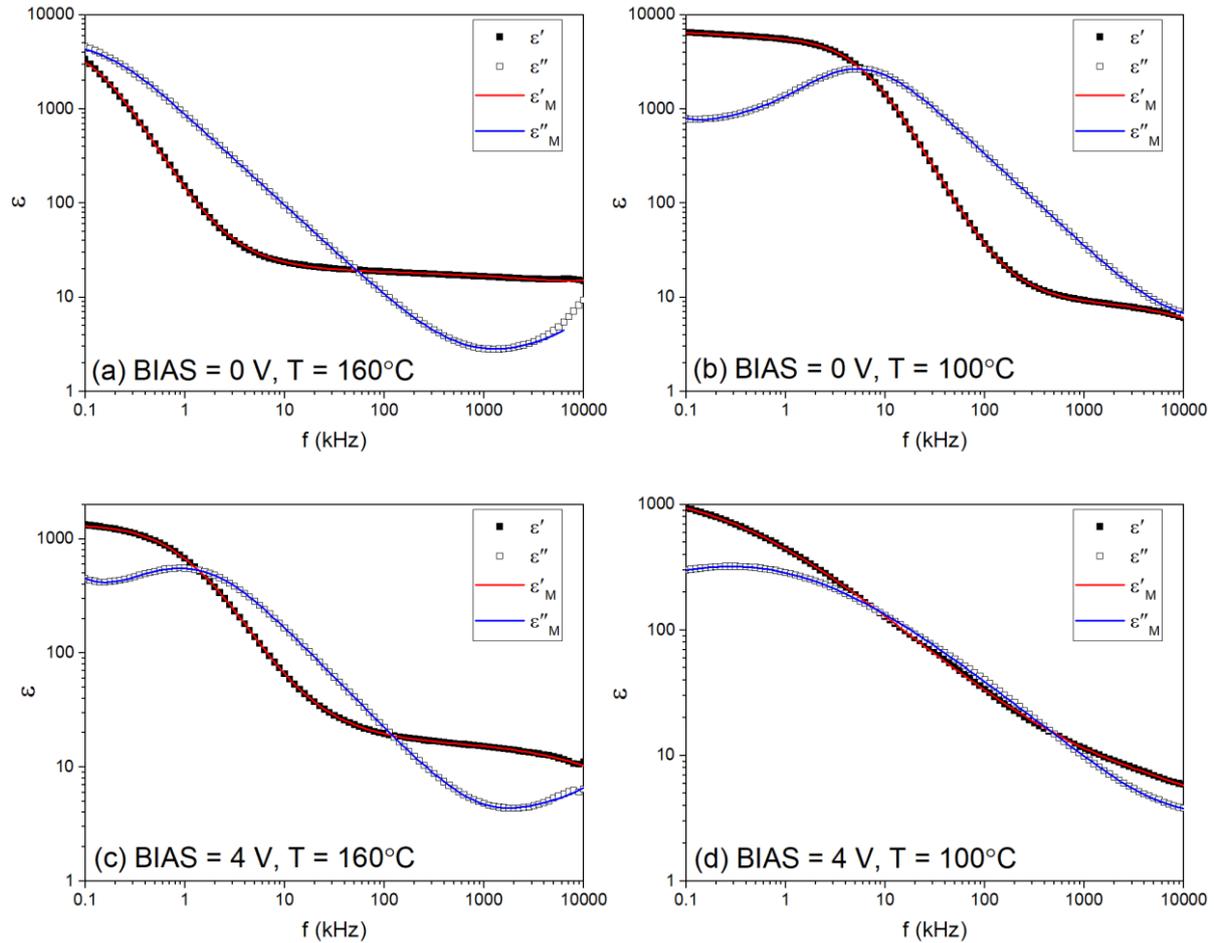

Figure 7 The example of fitting real $\varepsilon'$ (full black squares) and imaginary $\varepsilon''$ (empty black squares) part of the electric permittivity to the Cole-Cole model in the planar cell at (a, c) the nematic phase and (b, d) the ferronematic phase.

To understand the origin of the DC-created relaxation process visible in the N phase, named the Ferro-C, and the dielectric behavior of the Ferro-A and Ferro-B modes, we studied the effect of the bias electric field on the relaxation parameters. Figure 8 shows the temperature evolutions of the relaxation frequency $f_R$ and dielectric strength $\delta\varepsilon$. In the $N_F$ phase the ferroelectric modes A and B present a high dielectric strength ($\delta\varepsilon > 1000$) and a relatively low relaxation frequency ($f_R < 10$ kHz). We see that relaxation parameters and temperature dependence of ferroelectric modes are similar. For both modes, the relaxation frequency decreases with the temperature decreasing. It means that the modes are temperature-activated. The dielectric strength of Ferro-A is bigger than Ferro-B. Below 80°C $\delta\varepsilon$ of the ferroelectric modes significantly decrease with temperature decreasing. In the measurements, without the DC field, the ferroelectric modes are visible in the dielectric spectra to the temperature of ca. 130°C – the temperature range of the ferronematic phase. The low-frequency relaxation exists in the N, $N_F$, and Iso phases except in the region of 130°C -150°C. In this temperature region, the relaxation frequency drop below the frequency range of the impedance analyzer and it would be difficult to estimate the parameters of this relaxation. The low-frequency relaxation is responsible for the high value of electric permittivity in the nematic phase. This dielectric mode in the ferroelectric phase is weaker than the ferroelectric modes. Under the DC field of 4 V, the dielectric strength of the modes generally decreases. We no longer see two mods in the $N_F$ phase, but only one – Ferro-C, which was most likely a combination of modes Ferro-A and Ferro-B, but much weaker. The Ferro-C mode is also observed in the nematic phase. This mode was created by the applying DC field to the N phase and it is a continuation of the ferroelectric modes A and B. The low-frequency mode is still observed throughout the whole temperature range, but its relaxation frequency in the N phase is higher.

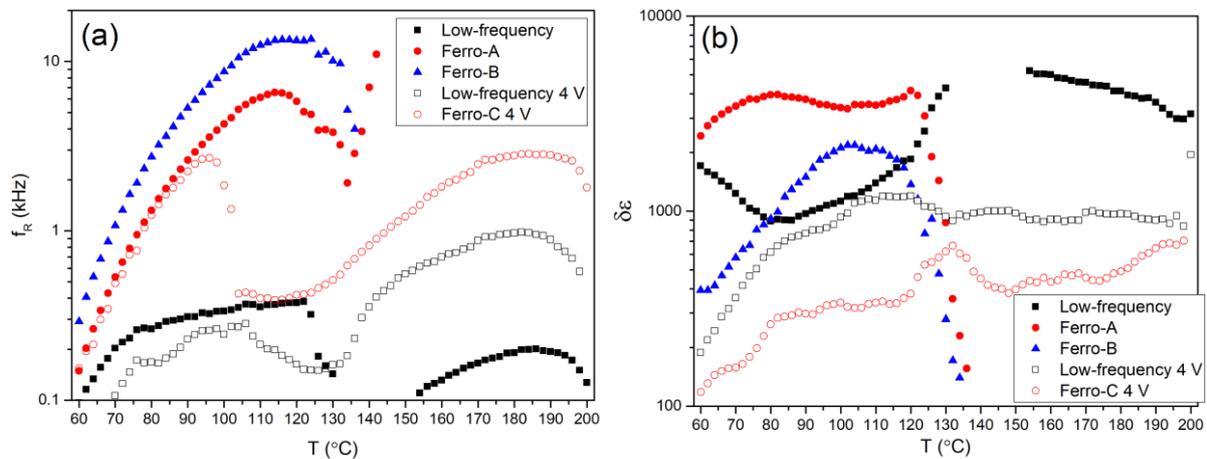

Figure 8 Temperature dependence of (a) $f_R$ relaxation frequency and (b) $\delta\varepsilon$ dielectric strength of observed dielectric modes in the ferronematic and nematic phase of RM734.

The ferroelectric response in DS controlled by the DC field, unlike an AC field, depends on polarity (Figure 9). After applying the DC field of positive polarity, the spontaneous polarization reorients towards one direction and the director reorients to the homeotropic alignment. This causes the latching of the ferroelectric state.[34] The latching of the ferroelectric state prevents the change of molecular orientation between the $+\vec{P}$ and $-\vec{P}$ state through the formation and movement of the ferroelectric domains. As a result, as the bias electric field strength increases, the value of the measured electrical permittivity decreases in the ferronematic phase. The low AC field (100 mV), generated by the impedance analyzer, cannot repolarize the $N_F$ phase. With the increasing bias field, the volume of frozen dipoles increases as long as all dipoles will be aligned along the bias electric field. After removing the DC field the director $\hat{n}$, and polarization vector $\vec{P}$ relaxes to the sinusoidal electrical signal profile generated by the impedance analyzer. At higher fields the coupling between $\hat{n}$ and $\vec{P}$ is much stronger than the dielectric effect and the time required to relax to the planar state is longer than for lower fields. On the other hand, from the C-C model, we can conclude that the application of the bias electric field results in the creation of the ferroelectric order in the nematic phase, which causes the increase of $\varepsilon'$ in the N phase. The confirmation of the phenomenon is the DC-created ferroelectric type relaxation. It can be considered as the similar effect to the electroclinic effect at the SmA*-SmC* phase transition.[35–37] In the dielectric studies the most pronounced increase of $\varepsilon'$ is observed at 2 V. For higher fields the effect associated with the latching of the ferroelectric state begins to play a leading role. Therefore, the increase of $\varepsilon'$ for 10 V is smaller than for 2 V. Another explanation for changing the dielectric constant in the material under the bias electric field is a creation of complex molecular objects. In highly polar isotropic liquids the bias field strongly modulated the value of electric permittivity in the isotropic phase. On the one hand, the increase of $\varepsilon'$ can be explained by the formation of dimers.[38] On the other hand, the decrease of the dielectric constant in the highly polar isotropic liquids is related to the formation of quadrupoles.[39]

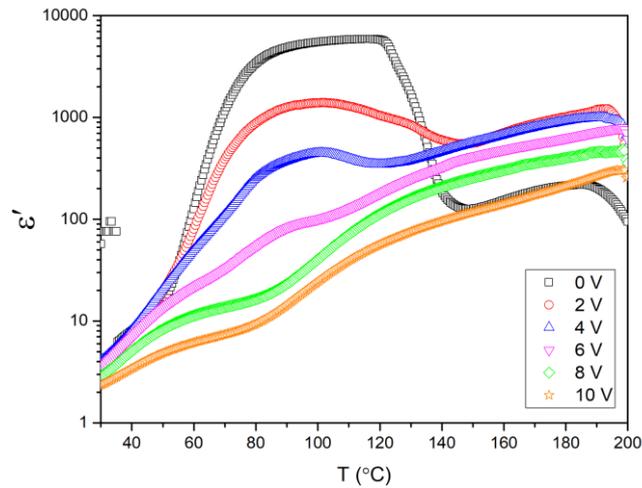

Figure 9 Real part of electric permittivity $\varepsilon'$ upon temperature $T$, for several DC fields (from 0 V to 10 V).

By measuring the spontaneous polarization using the AC signal we show that without the DC field, there is no ferroelectricity in the nematic phase. The existence of spontaneous polarization $P_s$ is the main feature of the $N_F$ phase. The $P_s$ was determined using the field reversal technique.[40] To obtain the spontaneous polarization-reversal current profile with a visible peak, the alternating triangular electric signal of amplitude $U = 40$ Vpp and frequency of $f = 5$ Hz was applied to the HT cell of thickness $d = 5.0 \pm 0.1$ μm. The values of $U$ and $f$ were chosen to observe the highest peak possible (Figure 10a). For higher frequencies than 5 Hz the peak was smaller and was not well-detectable. The spontaneous polarization was measured in the homeotropically aligned cell due the fact that the polarization vector is parallel to the liquid crystal director. In the electrical response the typical ferroelectric switching current was observed. From the area under the polarization current curve the value of $P_s$ was calculated. The observed $P_s$ increases initially with decreasing temperature. The polarization peak is started to be observed below 140°C, where the N-$N_F$ starts. The measured polarization of reach maximum of ~2.7 μC/cm² at 73°C and then decreases drastically. The value of $P_s = $ ~2.7 μC/cm² is slightly lower than to the value reported in the literature[17]. The decrease in spontaneous polarization is most likely related to an increase in viscosity. Below 60°C $P_s$ became immeasurable.

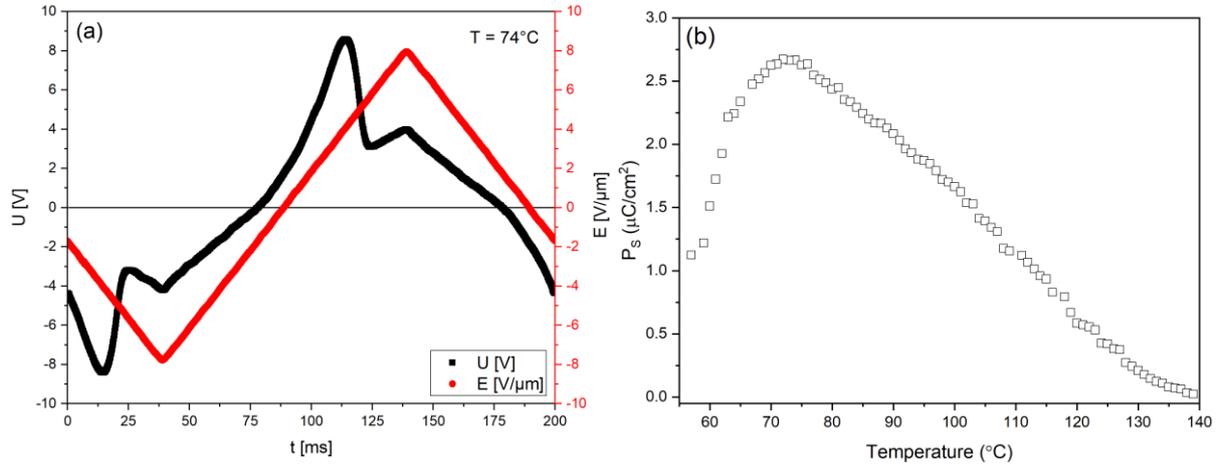

Figure 10 (a) Electrical response (black line) of the cell with the visible peak of the reverse polarity, after applying the triangular wave-shaped signal (red line). (b) Temperature dependence of the spontaneous polarization.

In a classical nematic liquid crystal, we can observe two types of molecular relaxation processes around two main axes: rotation around short (S-mode) and long (L-mode) molecular axes [41]. In the planarly aligned cell, the S-mode and the L-mode are usually not detectable by DS at room temperature, because of the high-frequency domain of detection, hundreds of MHz or a few GHz. However, cooling the sample below room temperature or having imperfect planar alignment exposes the relaxation spectra which can be related to the L or S dielectric mode. In ferroelectric nematic liquid crystals, the molecular system seems to be much more complicated. Due to the molecular packing of the $N_F$ phase, we propose the model where one of the ferroelectric relaxations behaves as a collective mode. We assume that within one domain the molecules vibrate in phase. While in the second, with opposite polarity, they move against the phase. The second ferroelectric mode can be related to some intrinsic degrees of freedom of a molecule bound to a polar group. The mode may be an analogy to the S-type mode. The collective mode should be characterized by higher dielectric strength and lower relaxation frequency than the molecular mode. Taking the above into consideration, we can assume that the Ferro-A mode can be described as a collective dielectric mode, while Ferro-B is a molecular mode. Their occurrence is closely related to the presence of the ferroelectric phase. Whereas in the nematic phase, the relaxation process is purely molecular. The molecules rotate individually in a chaotic manner. After applying the DC field, the collective mode, due to the latching of the ferroelectric state, is not visible in the dielectric spectra. For this reason, under the bias field, only one mode in the dielectric spectra is observed. This visible mode is probably the molecular mode. The ferroelectric molecular mode is observed also in the nematic phase, because of the generating the ferroelectric order and the switching of the molecular director to the homeotropic state. In the electro-optical studies, the DC-created ferroelectric order may be visualized, after applying the DC signal to the IPS cell, as domains with higher orientational order in Figures 5e-f. To optically confirm the creation of the ferroelectric order, a comprehensive study of induced birefringence would be required, but this topic is not the subject of the article. The low-frequency relaxation is most likely connected to ionic conductivity. This mode is suppressed under the influence of the bias electric field. Therefore, its dielectric strength decreases. Moreover, the low-frequency mode behaves differently in N and $N_F$ phases. In the ferroelectric phase where the molecules are in the domains, the ionic conductivity significantly decreases, and the dielectric strength also decreases, compared to the N phase.

**Conclusions**

In conclusion, we confirmed the existence of a new class of ferroelectric fluid. We showed the effective procedure for creating well-created lens-shaped planarly oriented domains of $N_F$ with opposite polarity. The rotation of the polarization vector in the ferroelectric nematic domains can be controlled simply by applying the AC triangular wave-shaped electric field or the DC field in the in-plane switching cell. The domain structure of the ferroelectric phase entails different behavior of molecular relaxation. In the ferronematic phase, two ferroelectric relaxations were found. We proposed the model where one relaxation exhibits a collective nature, while the second dielectric mode behaves as a typical molecular relaxation. Under the bias electric field, two phenomena were observed. The latching of the ferroelectricity state was revealed in the dielectric spectra by decreasing electric permittivity. While the generation of the ferroelectric order in the nematic phase was observed by detecting the weak ferroelectric mode in the N phase and the observation of ferroelectric domains in the IPS cell under the DC field (Figures 5e-f). The temperature dependence of the spontaneous polarization is characterized by a maximum of ~2.7 $\mu C/cm^2$ at a fixed temperature. Our results will find application in nonlinear optics[42] and new photonic devices based on the polarization steering phenomena.

**Data availability**

Data for this article are available in Zenodo at https://doi.org/10.5281/zenodo.10371947.

**Acknowledgments**

This work was supported by NCN MINIATURA project 2022/06/X/ST5/01316 and NCN OPUS project 2019/33/B/ST5/02658 and University project UGB 22-804 and 22-801.

**Author contribution**

MM conceived the project, performed POM, electro-optical studies, dielectric spectroscopy spontaneous polarization measurements, analyze, and wrote the manuscript with the inputs from all co-authors. PP analyzes the data. JK and PK synthesized the material and performed DSC studies. All authors contributed to scientific discussions.